\documentclass[12pt]{article}%
\usepackage[latin1]{inputenc}
\usepackage{amssymb}
\usepackage{amsmath}
\usepackage{latexsym}
\usepackage{amsthm}
\usepackage{eucal}
\usepackage{indentfirst}
\usepackage{epsfig}
\usepackage{amsfonts}
\usepackage{natbib}

\begin{document}

\title{A Fast Reconstruction Algorithm \\for Gene Networks}
\author{Lorenzo Farina and Ilaria Mogno\\\textit{Dipartimento di Informatica e Sistemistica "A. Ruberti"} \\\textit{Universit\`{a} degli Studi di Roma ''La Sapienza''} \\\textit{Via Eudossiana 18, 00184 Roma, Italy} \\\textit{e-mail: lorenzo.farina@uniroma1.it, mogno@dis.uniroma1.it} \\\textit{tel: +39-0644585690, +39-0644585938} \\\textit{fax: +39-0644585367}}
\maketitle

\begin{abstract}
This paper deals with gene networks whose dynamics is assumed to be generated
by a continuous--time, linear, time invariant, finite dimensional system
(\textit{LTI}) at steady state. In particular, we deal with the problem of
network reconstruction in the typical practical situation in which the number
of available data is largely insufficient to uniquely determine the network.
In order to try to remove this ambiguity, we will exploit the biologically
\textit{a priori} assumption of network sparseness, and propose a new
algorithm for network reconstruction having a very low computational
complexity (linear in the number of genes) so to be able to deal also with
very large networks (say, thousands of genes). Its performances are also
tested both on artificial data (generated with linear models) and on real data
obtained by \textit{Gardner et al.} from the SOS pathway in
\textit{Escherichia coli}.

\end{abstract}


\section{Introduction}

In this paper we will consider the most simple -- not trivial dynamical model
of a gene network, define a \textit{reverse engineering problem\footnote{In
the system and control community, this problem is commonly defined as an
\textit{identification} problem. We prefer here to adopt the terminology used
in the biological literature.}} and propose a fast algorithm to tackle this
problem. We will deal then with continuous--time, linear, time invariant,
finite dimensional system (\textit{LTI systems}) at steady state, in view of
the relevant biological literature (see, for example \cite{Yeung99:2002}).
Even though this oversimplified model may not be very realistic "far" from
steady state, nevertheless it is a fundamental tool for studying and gaining
insight into the basic mechanism that makes this problem an hard one, thus
providing a valuable \textit{in numero} testbed for gene networks
reconstruction algorithms. In fact, starting from the simplest and then moving
toward the more and more complex is a typical scenario in the applied
sciences. Consequently, in this paper we will consider only experiments
regarding steady state measurements due to "small" input constant
perturbations, so to retain linearity of the model.

More precisely, we deal with the problem of reconstructing a gene network in
the typical practical situation in which the number of available data is
largely insufficient to uniquely determine the network. In order to try to
remove this ambiguity, we will exploit some additional biologically relevant
\textit{a priori} assumptions such as sparseness, as it will be expalined in
the following.

The proposed algorithm has the major advantage to be very fast. In fact, its
complexity is $O(2NM^{2})$ and consequently it is particularly useful for
large networks (large $N$) and few data (small $M$).

Finally we will also generate artificial data for the reconstruction
algorithm. By doing so, one exactly knows the "true" gene regulation network
and then algorithms performance can be evaluated. Moreover, the algorithm will
be successfully applied to real data collected from experiments regarding a
nine-gene subnetwork of the SOS pathway in \textit{Escherichia coli} and very
recently presented in \cite{Gardner2003}.

\section{Dynamic model of expression data and problem formulation}

The idea of modelling gene expression data with differential equations has
been explored by a considerable numbers of authors (see for example references
\cite{Chen4:1999}, \cite{Daeseleer1999}, \cite{Yeung99:2002},
\cite{Holter98:2001}, \cite{Someren2000}, \cite{Weaver1999},
\cite{Dehoon:2003} and, for a literature review, the interested reader may
refer to \cite{DeJong9:2002}). Differential equations are used to model gene
interactions under the assumption that the rate of change over time of each
gene expression level is a function of the expression level of some (usually a
few) other genes. Such modeling assumption is based on the reaction kinetics
at the biochemical level. A fully realistic model should consider a number of
relevant biological issues such as, for example, internal and external noise,
time delays, specific classes of nonlinearities, and should also consider the
relationships between mRNA and protein concentrations since only the first one
is actually measured by microarrays. Clearly, a lot of work remains to be done
in the field of gene interaction modelling. Nevertheless, a very simple linear
time invariant model has proved to be useful in a number of cases (as in
\cite{Daeseleer1999} for rat cervical spinal cord data) even if it is clear
that nonlinearity is an unavoidable issue since it reflects also the nature of
biochemical interactions. However, as in \cite{Gardner2003}, we will assume
the system to behave linearly around the steady states points reached in
experiments. According to the work of Yeung \textit{et al.}
\cite{Yeung99:2002} we consider the LTI system described by the following
differential equations:
\begin{equation}
\dot{x}_{i}\left(  t\right)  =-\lambda_{i}x_{i}\left(  t\right)  +\sum
_{j=1}^{N}W_{ij}x_{j}\left(  t\right)  +b_{i}\left(  t\right)  +\xi_{i}\left(
t\right)  \label{MainEq}%
\end{equation}
for $i=1,2,...,N$, where the state variables $x_{i}$'s are the concentration
of mRNA measured as a difference from the equilibrium state preceding the
stimulus\footnote{Clearly, these numbers can be positive, negative or zero
provided that the measured concentration levels are greater, less or equal to
the concentration present prior to the stimulus.}, the $\lambda_{i}$'s are the
self--degradation rates, the $b_{i}$'s are the external stimuli (depending on
the specific experiment performed), and the $\xi_{i}$'s represent (internal)
noise. The elements of the matrix $W$ describe the type and strength of the
"influence" of the $j$-th gene on the $i$-th gene with a positive, zero or
negative sign indicating activation, no interaction and repression
respectively. As described in \cite{Gardner2003}, an experiment consists in
applying a prescribed (\textit{i.e.} known) stimulus $b_{i}\left(  t\right)  $
which is a persistent perturbation (ideally a step function $\delta_{1}(t)$ of
"amplitude" $b_{i}$, \textit{i.e.} $b_{i}\left(  t\right)  =b_{i}\delta
_{1}(t)$) to $M$ input separately and then use a microarray to measure the
response at steady state. Finally, the above situation, in view of equations
(\ref{MainEq}) and taking into account the measurement noise $w$ introduced by
the microarray device, can be formally described in the usual compact form as
follows:
\begin{align}
\dot{x}  &  =Ax+B+\xi\label{MainComp}\\
y  &  =x+w\nonumber
\end{align}
where the matrix $A$ is a $N\times N$ matrix which incorporates both
self--degradation rates (on its main diagonal entries) and the strength of the
gene--to--gene interaction (on its off diagonal entries) and the columns of
the $N\times M$ matrix $B$ are the $b_{i}$'s. We will assume standard
normality properties on zero mean noises.

Steady state (equilibria) solutions are given by $Ax_{e}+Bu_{e}+\xi
=0=Ay_{e}+Bu_{e}+\xi-Aw$, and if we repeat the $M$ measures a sufficiently
large number of times, we can calculate average values and then solve
$A\overline{y}+B\overline{u}=0$, where $E[y_{e}]=\overline{y}$, $E[u_{e}%
]=\overline{u}$ and $E[\xi-Aw]=0$ so that $E[w]=E[\xi]=0$.

We can now state the problem considered in
\cite{Gardner2003}:
\bigskip

\begin{center}
\textsc{An Identification (Reverse Engineering)}

\textsc{Problem for Gene Networks}
\end{center}

\textit{Given a network described by (\ref{MainComp}), initially
at rest composed of }$N$ \textit{genes, where }$M$\textit{\
independent constant inputs\footnote{The matrix $B$ is known and
full rank, possibly diagonal.} are applied and }$M$\textit{\ noisy
measurements }$y^{j}$\textit{\ are taken at steady state. This
process is repeated a "sufficiently" large number of times so that
we can average the data and obtain the data matrix }
$\bar{Y}_{N\times M}$\textit{, whose generic element is
}$\bar{y}_{i}^{j}$\textit{\ where superscripts denote experiments
(which are then repeated }$M$ \textit{times), the overbar denotes
averaging and the subscripts denote individual genes.}

\textit{Then, a \textbf{reverse engineering problem} consists in finding the
"best" matrix }$A$\textit{\ such that}
\begin{equation}
A_{N\times N}\bar{Y}_{N\times M}+B_{N\times M}=0_{N\times M} \label{MainComp2}%
\end{equation}

It is important to note that the matrix $A$ in equation (\ref{MainComp2}) is
unique if and only if $M\geq N$, \textit{i.e.} provided that the number of
experiments is equal or greater than the number of genes in the network. In
what follows we will assume the typical situation in which $M\ll N$ so that
equation (\ref{MainComp2}) is underdetermined, that is, it has many solutions
yielding $A\bar{Y}_{N\times M}+B=0$. Thus, in order to find the \textit{best
}choice for the matrix $A$, some \textit{a priori} information has to be
exploited incorporating some biological knowledge into the model at hand. One
possibility, as discussed in a previous section, is to try to impose the
additional biological constraint that usually gene networks are sparse,
\textit{i.e.} that generally each gene interacts with only a small percentage
of all the genes in the entire genome (see for example \cite{Jeong407:2000}).
We will therefore exploit this feature and present our algorithm in the
following section.

In the next paragraph, we will first address the problem of generating
artificial data consistent with the above biological assumption so to be able,
after presenting a new reverse engineering algorithm, to evaluate its
performances on such generated artificial data and, afterwards, also on real
data taken from \cite{Gardner2003} regarding the SOS pathway of
\textit{Escherichia coli}.

\section{The artificial data generation}

In this section we will briefly describe how one can set up an artificial
problem and the data generated by this procedure. First of all we generated
data assuming a linear model (as in (\ref{MainEq})) and the algorithms were
tested on such data. However, in this preliminary work we just wanted to
evaluate the performances of the algorithms presented in the next paragraphs.
We would like also to stress that we applied our reconstruction algorithms to
artificial data, before using real data. The reason simply being that, by
doing so, one exactly knows the "true" regulation network, and the algorithm
performances can be evaluated.

Consequently, in our first artificial experiments, we generate the data with
the linear model described in equation (\ref{MainEq}). Each repetition of the
$i$--th experiment consists, in this model, in applying a constant unitary
input (normalized) and then in picking a sample of the trajectory generated by
equations (\ref{MainEq}) at steady state. We generate then a \emph{sparse},
\emph{asymptotically stable} matrix $A$, and, as a second step, also the
entries of input matrix $B$ for each experiment. Finally, we obtain the
artificial data by considering steady state values, and by repeating this
process $M$ times, we get the simulated data matrix $\bar{Y}_{N\times M}%
$\footnote{As previously stated, the expression level of the genes are
calculated as the difference between the current expression level (the
cellular concentration of the gene product).}.

\section{A reconstruction algorithm for sparse gene networks}

As discussed in the previous sections, the reverse engineering problem for
gene networks when the number of measurement is (much) less than the number of
genes involved in the network, leads to many alternative networks among which
we have to find a way to choosing the "best" one according to a prespecified
cost function.

As stated by the reverse engineering problem, once we have collected the mRNA
abundance steady state measurements $\bar{Y}_{N\times M}$ from microarrays
with $M<N$, we may first find one of the many solutions $A$ of $A\bar{Y}+B=0$.
A standard procedure for the case of interest with $M<N$, is to perform a
\textit{singular value decomposition} on the data matrix $\bar{Y}_{N\times M}$
and obtain the matrix $A$ with the smallest $L_{2}$ norm. The procedure goes
as follows. We decomposed the data matrix $\bar{Y}_{N\times M} $\ as $\bar
{Y}_{N\times M}=U_{N\times N}S_{N\times M}W_{M\times M}^{T}$, where
$U_{N\times N}$ and $W_{M\times M}^{T}$ are unitary matrices and the entries
$\sigma_{ij}$ of the matrix $S_{N\times M}$\ are such that $\sigma_{ij}=0$ for
all $i\neq j$ and $\sigma_{11}\geq\sigma_{22}\geq\ldots\geq\sigma_{MM}>0$. The
numbers $\sigma_{ii}:=\sigma_{i}$ are the nonnegative square roots of the
eigenvalues of $Y_{N\times M}Y_{N\times M}^{T}$ known as the \textit{singular
values} of $Y_{N\times M}$. As previously stated, the SVD provide a simple way
to find a solution $A$ to problem (\ref{MainComp2}) with minimal $L_{2}$ norm,
\textit{i.e. }with minimal $\left\Vert A\right\Vert _{2}$, as\footnote{Note
that, in this case, we have $A_{svd}\bar{Y}_{N\times M}+B=0$.} $A_{svd}%
=-BWdiag_{i=1,\ldots,M}(\frac{1}{\sigma_{i}})U^{T}$. All the solutions to
(\ref{MainComp2})\ are given by $A=A_{svd}+C$,where $C$ is any $N\times N$
matrix satisfying $C_{N\times M}\bar{Y}_{N\times M}=0$, as one can easily
verify by direct substitution. Equivalently, the matrix $C$ is such that the
columns of $C^{T}$ belong to the null space of $\bar{Y}_{N\times M}^{T}$.

A possible way of "reasonably" choosing the "best" matrix $A$ without
evaluating all the feasible solutions, is given next. Our aim is to take into
account the sparseness assumption, that is try to incorporate information on
the structure (zero pattern) of the matrix $A$. In particular, we will
concentrate on the case in which the number $k$ of nonzero entries in each row
of $A$ is not greater than the number of available measurements, \textit{i.e.}
$k\leq M$. Consequently, we have that $k\leq M<N$ with $k\ll N$. The proposed
procedure, hereafter formally described, consists of three steps:

\begin{center}
\textsc{The Fast Reverse Engineering Algorithm}
\end{center}

\begin{enumerate}
\item Assuming $k\leq M<N$ and using the full rank data matrix $\bar
{Y}_{N\times M}$, find the minimal $L_{2}$ norm solution $A_{svd}%
=-BWdiag_{i=1,\ldots,M}(\frac{1}{\sigma_{i}})U^{T}$ to equation $A\bar
{Y}_{N\times M}+B=0 $ using SVD of the data matrix $Y_{N\times M}$.

\item For each of the $N$ rows of $A_{svd}$, determine the smallest in
magnitude $N-M+\eta$ with $\eta\geq1$ entries and consider them as "zero".

\item Taking into account the zero pattern detected at the previous step, find
row by row the unique (sparse) matrix $A$ solution of $\min\arg\left\Vert
A\bar{Y}_{N\times M}+B\right\Vert $, using the pseudo-inverse operator on the
selected submatrix, and we are done.\bigskip
\end{enumerate}

Here, the basic idea is to exploit the fact that, at the first step, the
computed matrix $A_{svd}$ is that of minimal $L_{2}$ norm, so that it is
\textit{reasonable} to assume that the smallest magnitude entries correspond
to zeros in the "true" matrix $A$. The main advantage of this approach is its
simplicity and low computational complexity as it will discussed later in the
paper. It's worth noting that Step 3 requires only that, for each of the $N$
rows of $A$, an $(M-\eta)\times M$ matrix to be pseudo-inverted and this can
be done in a very efficient computational way. The complexity of the algorithm
is then $O(2NM^{2})$ so it is particularly useful for large networks (large
$N$) and few data (small $M$).

We compare here this algorithm with another one which makes use of an
heuristics to find the zero pattern for each row of the matrix. Instead of
applying Step 3 (of the \emph{Fast} algorithm), for each row $i$, we put to
zero one entry, say $a_{ij}$, of the matrix $A_{svd}$ and denote this modified
matrix by $A_{svd,ij}$. We evaluate then the error index $\delta_{ij}=\Vert
A_{svd,ij}Y+B\Vert_{2}$. The position of the first zero in the $i$-th row is
then that corresponding to the minimal value of $\delta_{ij}$. We repeat this
process until we have placed all the $N-M+\eta$ zeros. More precisely, using
this criterion, we find for each row of the matrix $A_{svd}$ the $N-M+\eta$
zero locations and consequently we can solve uniquely the system by selecting
the $M-\eta$ nonzero elements for each row $i$.

Simulation results are provided and discussed in the following section.

\section{Results on artificial data}

In this section, we will present the simulation results and the performances
of the proposed reconstruction algorithm. First of all, we generate artificial
data as described in the previous sections, then we estimate the matrix $A$,
applying the reconstruction algorithms with $\eta=1$. We measure the error by
counting the percentage of sign discrepancies\footnote{The information on the
sign of the entries is the most relevant information from a biological point
of view since reflect the nature of the gene-to-gene interaction.}
$E=100\frac{\sum_{i=1}^{N}\sum_{j=1}^{N}e_{ij}}{N^{2}}$ where
\[
e_{ij}=\left\{
\begin{array}
[c]{l}%
1\qquad\text{if sign}\left(  a_{ij}\right)  =\text{sign}\left(  \bar{a}%
_{ij}\right)  \\
0\qquad\text{otherwise}%
\end{array}
\right.
\]%

\begin{figure*}[t]
\centerline{\leavevmode \epsfysize=4cm \epsfbox{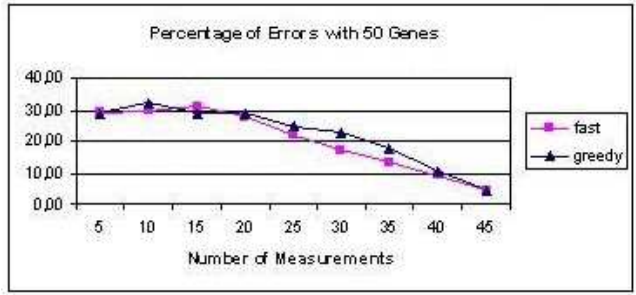}}
\centering \caption{\small{The percentage of errors estimated by
the two algorithms vs. the number of available measurements (case
of 50 genes). Data are generated with the linear model.}}
\label{fig1}
\end{figure*}

In Figure \ref{fig1} we show the simulation results of a network composed by
$50$ genes. We have generated the data using the linear model. In the figure
it is clear how both the algorithms we propose produce good results.%

\begin{figure*}[t]
\centerline{\leavevmode \epsfysize=4cm \epsfbox{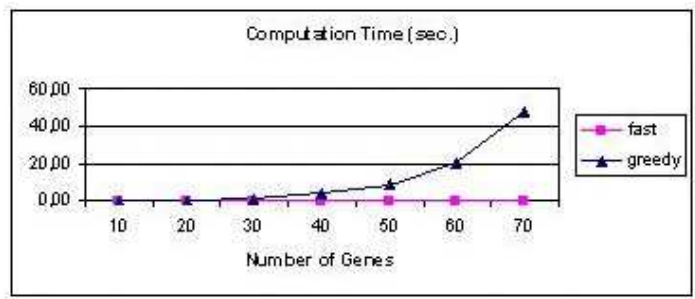}}
\centering \caption{\small{The computing time of the two
algorithms vs. the number of genes in the network with $M=N/2$.}}
\label{fig2}
\end{figure*}

In Figure \ref{fig2} we show the computing time\footnote{The algorithms
described here have been fully implemented with Matlab. They ran on a Pentium
IV with a clock speed of 1.8 GHz} of the two algorithms. It is clear that the
proposed \emph{Fast} algorithm grows much less than \emph{Greedy} search. If
the number of the genes of the analyzed network increases (say $200$, $500$,
$1000$ or more genes) any exhaustive or greedy search, takes to much time to
be computed. This is the strength of the algorithm we propose here.

\section{Results on real data}

As an example of application of the algorithm proposed in this paper, we
consider the data obtained in \cite{Gardner2003} for a nine-gene subnetwork of
the SOS pathway in \textit{Escherichia coli}. The collected data, \textit{i.e.
}the diagonal input matrix $B_{9\times9}$ and the data matrix $Y_{9\times9}$,
is reported in the web supplement of the
paper\footnote{http://www.sciencemag.org/cgi/content/full/301/5629/102/DC1}.%

\begin{figure*}[t]
\centerline{\leavevmode \epsfysize=8cm \epsfbox{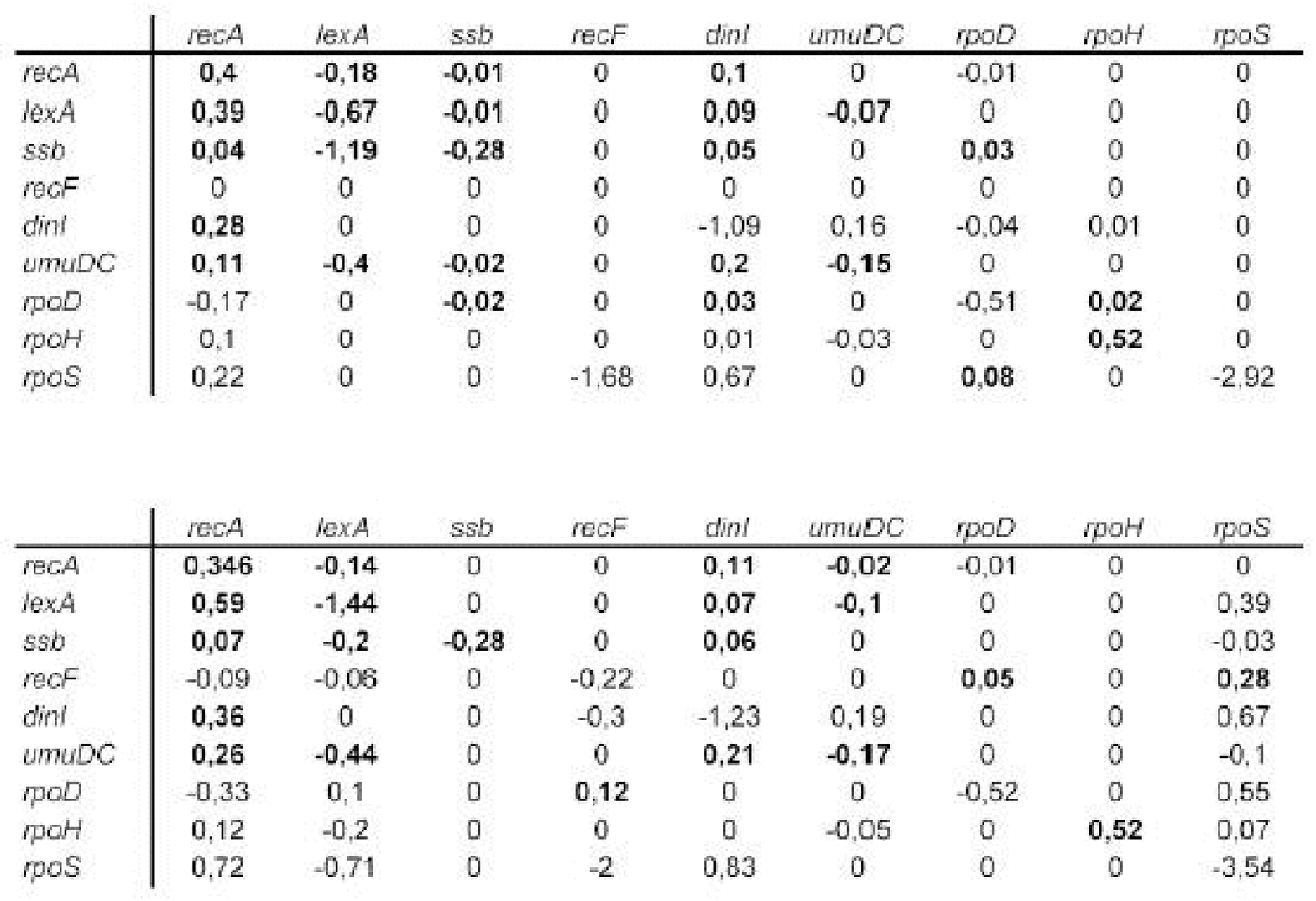}}
\centering \caption{\small{The table (labelled with gene names) in
the upper part of the picture is the identified matrix by Gardner
et al., those in the lower part is obtained using our method.
Numbers in bold indicate a correct\ sign. Note that a positive
sign on the main diagonal denotes positive self-regulation apart
from the self degradation rate, so that stability is guaranteed.
The recF row must be disregarded since the experiments didn't
yield reliable data.}} \label{fig3}
\end{figure*}

It is apparent from the the table in Figure \ref{fig3}, that the estimated
matrix $A$ performs quite well since the numbers of correct gene interaction
signs (as reported by the biological literature on the SOS pathway in
\textit{E. coli}) is not very different from that of \textit{Gardner et al}.
It is very important to note that the approach used by \textit{Gardner et al.
}considers, for each row, all the possible positions of the $4$ zeros in the
dynamic matrix $A$, thus leading to a combinatorial approach since it
evaluates every feasible solution. This is viable only for small subnetworks,
while our method is suitable also for very large number of genes (\textit{e.g.
}the \textit{Saccaromyces Cerevisiae }has about 6000 genes!)\textit{\ }being
polynomial in the number of genes.

\section{Conclusions}

In this paper we have addressed the so called "reverse engineering" problem,
that is the problem of reconstructing the gene-to-gene interactions from
expression data (noisy samples of the state space trajectories) in the usual
case of "few" samples and "many" genes. The typical situation is $\sim10$
samples for gene networks with $\sim5000$ genes, as in \cite{Gash2000}.

We have proposed a \emph{fast} algorithm for finding a "reasonable" solutions
among the many feasible by exploiting the \textit{a priori }biological
information about sparseness of the gene networks. However, this hypothesis in
too generic for selecting a "small" set of solutions, but it is of great help.
The main problem is that we only know that there are "many" zeros, but no
assumptions are made about the locations of such zeros. This makes even the
problem of imposing sparseness a difficult problem since it has an intrinsic
combinatorial complexity. To avoid this, in the literature many approaches
have been developed mainly using a \textit{greedy }search strategy, but still,
the computation burden is very high. In this paper we have proposed and
evaluated on artificial and real data, an algorithm with a very low
computational complexity (linear in the number of genes) which can be used
also for genome wide measurements. From this point of view, the algorithm
presented seems very promising and future work will be devoted to explore its
potential on large scale real data.

\bibliographystyle{unsrt}
\bibliography{fast}

\end{document}